\documentclass[twocolumn]{aastex631}
\usepackage{booktabs}
\usepackage{amsmath}
\shorttitle{Unobscured QSOs with KS4}

\begin{document}

\title{Exploring Unobscured QSOs in the Southern Hemisphere with KS4 }

\correspondingauthor{Minjin Kim}
\email{yjkim.ast@gmail.com; mkim.astro@gmail.com}

\author[0000-0003-1647-3286]{Yongjung Kim}
\affiliation{Korea Astronomy and Space Science Institute, Daejeon 34055, Republic of Korea}

\author[0000-0002-3560-0781]{Minjin Kim}
\affiliation{Department of Astronomy and Atmospheric Sciences, College of Natural Sciences, Kyungpook National University, Daegu 41566, Republic of Korea}

\author[0000-0002-8537-6714]{Myungshin Im}
\affiliation{SNU Astronomy Research Center, Seoul National University, 1 Gwanak-ro, Gwanak-gu, Seoul 08826, Republic of Korea}
\affiliation{Astronomy Program, Department of Physics \& Astronomy, Seoul National University, 1 Gwanak-ro, Gwanak-gu, Seoul 08826, Republic of Korea}

\author[0000-0002-3118-8275]{Seo-Won Chang}
\affiliation{SNU Astronomy Research Center, Seoul National University, 1 Gwanak-ro, Gwanak-gu, Seoul 08826, Republic of Korea}
\affiliation{Astronomy Program, Department of Physics \& Astronomy, Seoul National University, 1 Gwanak-ro, Gwanak-gu, Seoul 08826, Republic of Korea}

\author[0009-0003-1280-0099]{Mankeun Jeong}
\affiliation{SNU Astronomy Research Center, Seoul National University, 1 Gwanak-ro, Gwanak-gu, Seoul 08826, Republic of Korea}
\affiliation{Astronomy Program, Department of Physics \& Astronomy, Seoul National University, 1 Gwanak-ro, Gwanak-gu, Seoul 08826, Republic of Korea}

\author[0000-0002-7762-7712]{Woowon Byun}
\affiliation{Korea Astronomy and Space Science Institute, Daejeon 34055, Republic of Korea}

\author[0000-0002-1294-168X]{Joonho Kim}
\affiliation{Daegu National Science Museum, 20, Techno-daero 6-gil, Yuga-myeon, Dalseong-gun, Daegu 43023, Republic of Korea}

\author[0000-0002-6925-4821]{Dohyeong Kim}
\affiliation{Department of Earth Sciences, Pusan National University, Busan 46241, Republic of Korea}

\author[0000-0002-4179-2628]{Hyunjin Shim}
\affiliation{Department of Earth Science Education, Kyungpook National University, Daegu 41566, Republic of Korea}

\author[0000-0002-4362-4070]{Hyunmi Song}
\affiliation{Department of Astronomy and Space Science, Chungnam National University, Daejeon 34134, Republic of Korea}




\begin{abstract}

We present a catalog of unobscured QSO candidates in the southern hemisphere from the early interim data of the KMTNet Synoptic Survey of Southern Sky (KS4). The KS4 data covers $\sim2500\,{\rm deg}^{2}$ sky area, reaching 5$\sigma$ detection limits of $\sim22.1$--22.7 AB mag in the $BVRI$ bands. Combining this with available infrared photometric data from the surveys covering the southern sky, we select the unobscured QSO candidates based on their colors and spectral energy distributions (SEDs) fitting results. The final catalog contains 72,964 unobscured QSO candidates, of which only 0.4\,\% are previously identified as QSOs based on spectroscopic observations. Our selection method achieves an 87\,\% recovery rate for spectroscopically confirmed bright QSOs at $z<2$ within the KS4 survey area. In addition, the number count of our candidates is comparable to that of spectroscopically confirmed QSOs from the Sloan Digital Sky Survey in the northern sky. These demonstrate that our approach is effective in searching for unobscured QSOs in the southern sky. Future spectro-photometric surveys covering the southern sky will enable us to discern their true nature and enhance our understanding of QSO populations in the southern hemisphere.
\end{abstract}


\keywords{Quasars (1319) --- Photometry (1234) --- Catalogs (205) --- Galaxies (573)}


\section{Introduction} \label{sec:intro}

Quasi-stellar objects (QSOs), also known as quasars, are the brightest population of active galactic nuclei (AGN), which are important for exploring the evolution of the supermassive black holes (SMBHs) and their impact on host galaxies along the cosmic time. 
Unobscured (or type-1) QSOs, in particular, provide a unique opportunity to directly observe their central structure, including the accretion disk surrounded by dense ionized gas (broad line region: BLR) and dusty torus \citep{Urry95,Netzer15}.  Therefore, multiwavelength studies of those sources provide insights into detailed physical properties of AGNs (e.g., black hole mass and accretion rate, and thus the Eddington ratio; \citealt{Vestergaard06,Kelly09,Jun15,KimD23}) and structural parameters (e.g., physical sizes of the accretion disk, BLR, and torus; \citealt{Peterson04,Bentz09,Fausnaugh16,Cackett18,Cackett20,Lyu19,Yang20,KimM24,Woo24}).

Furthermore, there is a substantial interest not only in investigating the properties of individual objects but also in understanding the demography of QSOs. This includes statistical analysis of QSO populations, examining their distribution in space and time \citep{Hartwick90,Shen13}, and analyzing luminosity functions to trace cosmic evolution \citep{Hopkins07,Shen20,KimIm21}, alongside the corresponding black hole mass functions and distributions of Eddington ratio \citep{Kollmeier2006,Kelly13,Cho24}. Unobscured QSOs account for a significant portion of the total QSO population, making them valuable for such studies \cite[e.g.,][]{Treister2008}. Therefore, surveys of unobscured QSOs are essential, as they contribute to our understanding of both cosmic evolution and the mechanisms underlying the growth of SMBHs \cite[e.g.,][]{Yu2002, Martini2004}.

Despite significant advancements in QSO surveys over the past two decades (e.g., \citealt{Schneider02,Schneider10,Kim15,Kim19,Kim20,Kim22,Paris18,Lyke20,Onken23,Onken24}), the exploration of QSOs in the southern hemisphere remains incomplete. For example, \cite{Croom01} spectroscopically discovered $\sim 10$K QSOs in a field of a limited area  ($\sim$ a few hundred square degrees) in the southern hemisphere. \cite{Onken23} found 156 spectroscopically identified QSOs from the All-sky BRIght Complete Quasar Survey (AllBRICQS), based on the combination of All-sky space missions. In addition, \cite{Yang23} utilized optical photometric data ($grizY$) from the Dark Energy Survey (DES; $\sim 5,000\,{\rm deg^2}$) to search for the photometric QSO candidates in the southern hemisphere. 

Some studies have been more dedicated to exploring the particular area, the south ecliptic pole field, using optical imaging data \cite[e.g.][]{Byun23}. This region will be extensively explored by several survey missions, such as Euclid \citep{Euclid22}, Spectro-Photometer for the History of the Universe, Epoch of Reionization, and Ices Explorer (SPHEREx; \citealt{Dore14}), eROSITA \citep{Merloni12}, and 7-Dimensional Sky Survey (7DS; M. Im, 2024 in preparation). Therefore, to fully utilize the spectral and temporal survey data in the study of AGN physical properties, it is vital to pre-select the bright AGN in the southern hemisphere.

\begin{figure}
\centering
\epsscale{1.2}
\plotone{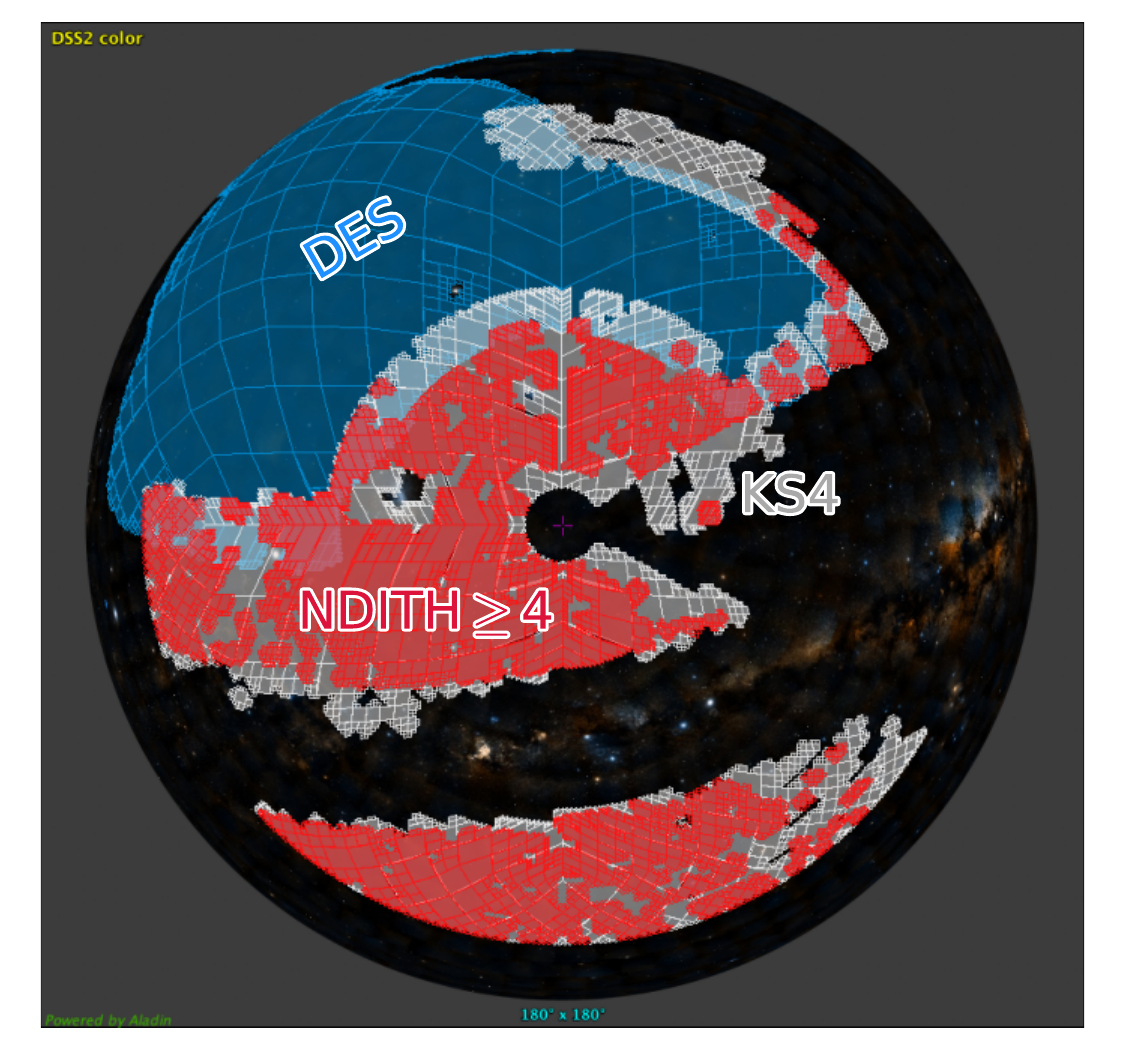}
\caption{Sky coverage map for the KS4 data, centered on the south ecliptic pole. Areas surveyed at least four times in the $BVRI$ bands using dithering techniques (\texttt{NDITH\_BVRI}$\,\geq4$) are marked in red. The DES survey area is shown in blue. The background represents a color composite from the Digital Sky Survey 2.
\label{fig:coveragemap}}
\end{figure}

Using a network of three KMTNet 1.6-m telescopes in Chile, Australia, and South Africa (\citealt{KimS16}), an optical imaging survey called the KMTNet Synoptic Survey of Southern Sky (KS4) is underway. This survey commenced on November 29, 2019. It aims to image up to 7000 square degrees of the sky visible from the southern hemisphere (${\rm Decl.} < -30^{\circ}$) in four optical bands (Johnson-Cousins $B, V, R$ and $I$), achieving 5$\sigma$ imaging depths of $\sim 22.1$--22.7 mag. It is noteworthy that the survey area has not been fully covered by DES. The primary science goal is to identify the optical counterparts of gravitational wave triggers and to study kilonovae in their early stages \citep{Kim2021ApJ...916...47K, Paek2024ApJ...960..113P}. We plan to roll out Data Release 1 in late 2024, which will be accompanied by a journal paper on the survey (Im et al. 2024, in preparation; Chang et al. 2024, in preparation) and its data processing methods (Jeong et al. 2024, in preparation). With its deeper imaging capabilities compared to other southern sky surveys, such as SkyMapper Southern Survey \citep{Onken24}, KS4 is expected to facilitate the discovery of more QSO candidates that were previously missed. The early interim data, which covers the sky area of $\sim2500\,{\rm deg}^{2}$ (Figure \ref{fig:coveragemap}), are currently available only to internal collaborators for research and verification.
Leveraging this early data, we aim to compile a catalog of unobscured QSO candidates, serving as groundwork for future survey missions.

The QSO selection solely based on the optical colors has some limitations due to contaminators, such as early type stars or star-forming galaxies. On the other hand, mid-infrared (MIR) colors are particularly effective in identifying QSO candidates \citep{Lacy04,Assef18}, largely because the MIR light is emitted from the hot or warm dust surrounding SMBHs, as posited by the unified AGN model \citep{Urry95,Netzer15}. However, since MIR wavelengths are less affected by dust extinction, relying solely on MIR colors can lead to a mixed sample of both unobscured and obscured QSOs.  The integration of optical data significantly enhances the selection of unobscured QSOs \citep{Byun23,KimD23,KimY24}. Therefore, we utilize a multi-wavelength dataset, including the KS4 data, to specifically target the identification of unobscured QSOs in the southern sky.

In this paper, we present the catalog of unobscured QSO candidates in the southern hemisphere, selected from the KS4 interim data. The KS4 data and photometry are described in Section \ref{sec:data}. In Section \ref{sec:selection}, we describe how the unobscured QSO candidates are selected, and the validation of the candidate selection is given in Section \ref{sec:validation}. Through this paper, we adopt the canonically used cosmological parameters for the standard $\Lambda$CDM universe: $H_{0}=70$ km s$^{-1}$ Mpc$^{-1}$, $\Omega_{m}=0.3$ and $\Omega_{\Lambda}=0.7$.  All the magnitudes are given in the AB system unless exceptions are noted.

\section{KS4 Data} \label{sec:data}
The data used in this work consists of object catalogs from $BVRI$ stacked images for 638 KS4 tiles ($\sim$2\,500 deg$^{2}$). Each pre-defined tile covers a total field of 2 degree $\times$ 2 degree, matching the field-of-view of the KMTNet camera \citep{KimS16}. Observations were made using three identical 1.6-m telescopes between late November 2019 and January 2022. We prioritized observations in areas of the sky excluding the Large Magellanic Cloud, Small Magellanic Cloud, low Galactic latitude regions ($|b| <$ 9 deg), and the regions already covered by the DES \citep{DES2016MNRAS.460.1270D, DESDR22021ApJS..255...20A}. Because of the altitude limit of KMTNet, the ${\rm Decl.} < -85^{\circ}$ regions are excluded as well. The basic survey strategy is straightforward: each tile is observed with a 120-second exposure per visit, with a total combined exposure time of at least 480 seconds. Each raw image consists of an array of four e2v CCDs (9k-by-9k array size, 10-$\mu$m pixel size) with a pixel scale of 0.4 arcsec. To fill the CCD gaps in the vertical (6.22 arcmin) and horizontal (3.07 arcmin) directions, we use a four-point dithering pattern with offsets of about 4 and 7 arcmin in RA and Dec, respectively. This dithering pattern results in heterogeneous image depths across the CCD gap and edge regions of each tile. 

The data were pre-processed at the KMTNet data center in Daejeon, Korea, including overscan and crosstalk corrections, bias subtraction, and flat fielding. A bad pixel mask is generated by identifying pixels that are affected by cosmic-ray hits on the detector, crosstalks from saturated pixels in an associated amplifier \citep{Kim06}, and regions that are unusable due to inherent defects in the CCD. The pre-processed data then undergoes zero-point (ZP) scaling to a uniform value of ZP = 30 mag, eliminating spatially variable ZPs across the field. For astrometric calibration, we utilize the \texttt{SCAMP} software \citep{Bertin2006ASPC..351..112B} to derive an accurate World Coordinate System (WCS) solution, using Gaia EDR3 \citep{Gaia16} as the reference catalog. This method achieves a root-mean-square error of 0.026 arcsec in both RA and Dec relative to the reference catalog WCS. 

The KS4 stacking procedure involves using the \texttt{SWarp} software \citep{2010ascl.soft10068B} to combine the pre-processed images. We employ a tangential projection method and a median combine technique to produce the final stacked images. Additionally, the \texttt{FSCALASTRO} parameter is set to \texttt{VARIABLE} to adjust for spatial variations in the pixel scale of KMTNet images, ensuring accurate flux rescaling throughout the field. The average seeing sizes of the combined images are about 2.10, 2.03, 1.94, and 1.93 arcsec in the $BVRI$ bands, respectively. 

The photometric calibration of the images is performed using APASS DR9 \citep{2016yCat.2336....0H} and SkyMapper DR3 \citep[SMSS;][]{2019PASA...36...33O} catalogs as reference data, with the following calibration equations: 
\begin{align*}
B &= B_{\rm APASS} - 0.06 - 0.27 \times (B_{\rm APASS} - V_{\rm APASS}) \\
V &= V_{\rm APASS} + 0.02 \\
R &= r_{\rm APASS} + 0.0383 - 0.3718 \times (r_{\rm APASS} - i_{\rm APASS}) \\
I &= i_{\rm SMSS} + 0.011 - 0.243 \times (i_{\rm SMSS} - z_{\rm SMSS})
\end{align*}

\noindent In the above, $B_{\rm APASS}$ and $V_{\rm APASS}$ are in Vega magnitudes, while the others are in the AB system. The references for these conversion equations are as follows: the $B$ and $V$ conversion equations are from \citet{2017ApJ...848...19P}, the $R$ is from \citet{2007AJ....133..734B}, and the $I$ is derived by comparing KMTNet images taken in the COSMOS field with the SMSS catalog of the same field. For point sources brighter than 17 mag, the average uncertainties of KS4 photometric calibration are 0.026, 0.023, 0.025, and 0.026 mag for the $BVRI$ bands, respectively. Correspondingly, the final 5$\sigma$ imaging depths are $B=22.75$, $V=22.60$, $R=22.80$, and $I=22.09$ mag, which are comparable to those of the preceding QSO survey with KMTNet data \citep{Byun23}. The saturation levels are 12.95, 13.07, 13.77, and 13.60 mag, respectively.

Source detection was performed on the KS4 $I$-band images with a signal-to-noise ratio criterion of \texttt{MAG\_AUTO\_SNR\_I}$>1$. Subsequently, we extracted the flux of sources in each band image using the dual mode of \texttt{SExtractor} \citep{Bertin96}. We primarily use the \texttt{MAG\_AUTO} estimates from the \texttt{SExtractor}, which measure flux within an adaptively scaled aperture. This approach is due to variations in the point spread function (PSF) depending on the source's position in the image.

To utilize infrared photometric data in the following analysis, we crossmatch the KS4 catalog with external catalogs from the infrared surveys covering the KS4 area: Two Micron All Sky Survey (2MASS; \citealt{Skrutskie06}), VISTA Hemisphere Survey (VHS DR5; \citealt{McMahon13}), and Wide-field Infrared Survey Explorer (WISE; \citealt{Wright10}). For each KS4 object, we identify potential counterparts by searching for the nearest neighbor source in external catalogs, relying solely on its coordinates. Considering the imaging resolution of the surveys, we tested several cases and established the matching radii for 2MASS, VHS, and WISE as 2.0, 1.0, and 2.0 arcsecs, respectively.

The 2MASS point source catalog (PSC) provides default magnitudes for the $J$, $H$, and $K_{s}$ bands (\texttt{j\_m}, \texttt{h\_m}, \texttt{k\_m}), representing the optimal measurements for each band. Although 2MASS is an all-sky survey, its detection limits are relatively shallow, with 5$\sigma$ detection limits for a point source at 17.4, 17.2, and 16.9 mag, respectively. To enhance our depth of detection, we also utilize $J$, $H$, and $K_s$ magnitudes from VHS, specifically the 2 arcsec diameter aperture magnitudes (\texttt{JAPERMAG3}, \texttt{HAPERMAG3}, \texttt{KSAPERMAG3}), which achieve 5$\sigma$ detection limits of 21.5, 21.2, and 20.3 mag, respectively. Note that VHS coverage in the $H$-band is limited to the south galactic cap. When a source appears in both 2MASS and VHS, we prioritize the VHS magnitudes, resorting to 2MASS only when VHS data is unavailable.

Among the various versions of WISE data, we use the W1, W2, W3, and W4 magnitudes from the AllWISE catalog \citep{Cutri13}. The magnitudes are measured with profile-fitting photometry (\texttt{w1mpro}, \texttt{w2mpro}, \texttt{w3mpro}, \texttt{w4mpro}) and their 5$\sigma$ detection limits are 19.6, 19.3, 16.7, and 14.6 mag, respectively.

To convert the Vega magnitudes to AB magnitudes, the following papers and conversion factors are referenced: \cite{Blanton05} for 2MASS $JHK_{s}$ bands (0.91, 1.39, 1.85), \cite{Gonzalez18} for VHS $JHK_{s}$ bands (0.916, 1.366, 1.827), and \cite{Cutri12} for WISE W1-W4 bands (2.699, 3.339, 5.174, 6.620).

For the final catalog, we perform the galactic extinction correction by utilizing \texttt{sfdmap} Python package\footnote{\url{https://github.com/kbarbary/sfdmap}}. Following the default setting, we use the dust map by \cite{Schlegel98} with a scaling factor of 0.86 from \cite{Schlafly11}. These corrections are made under the assumption of $R_V=3.1$.

\section{QSO Selection} \label{sec:selection}

\subsection{Photometric Selection} \label{sec:colorsel}

\begin{figure*}
\centering
\epsscale{1.2}
\plotone{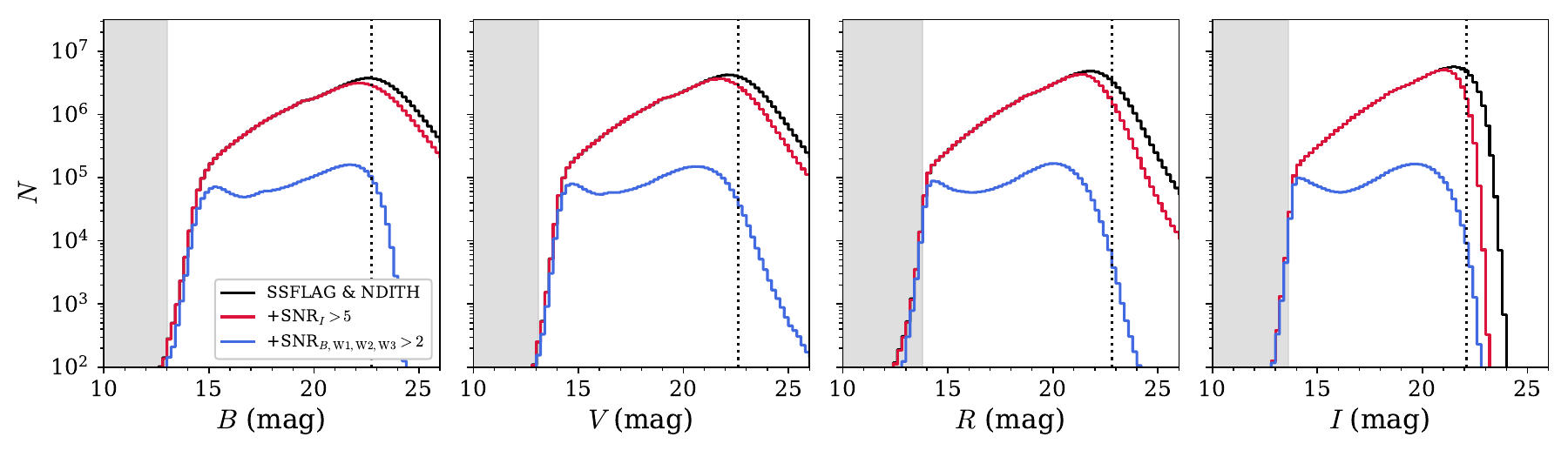}
\caption{Histograms of the reliable sources selected by \texttt{SSFLAG} and \texttt{NDITH\_BVRI} criteria across the $BVRI$ bands. The forced photometry in the $I$-band yields flux measurements of numerous sources beyond the 5$\sigma$ imaging depths, as indicated by the vertical dotted lines in each panel. The red histograms represent the marginal reduction in source count after applying an SNR criterion in the $I$-band. The blue histograms show the decrease in numbers when applying additional SNR criteria across the $B$/W1/W2/W3 bands. Note that the sudden increase in number at $I\lesssim16$ mag is due to the bright stars that remain after the SNR cut. The shaded regions denote the ranges of saturated magnitudes.
\label{fig:histdepth}}
\end{figure*}

\begin{figure*}
\centering
\epsscale{1.2}
\plotone{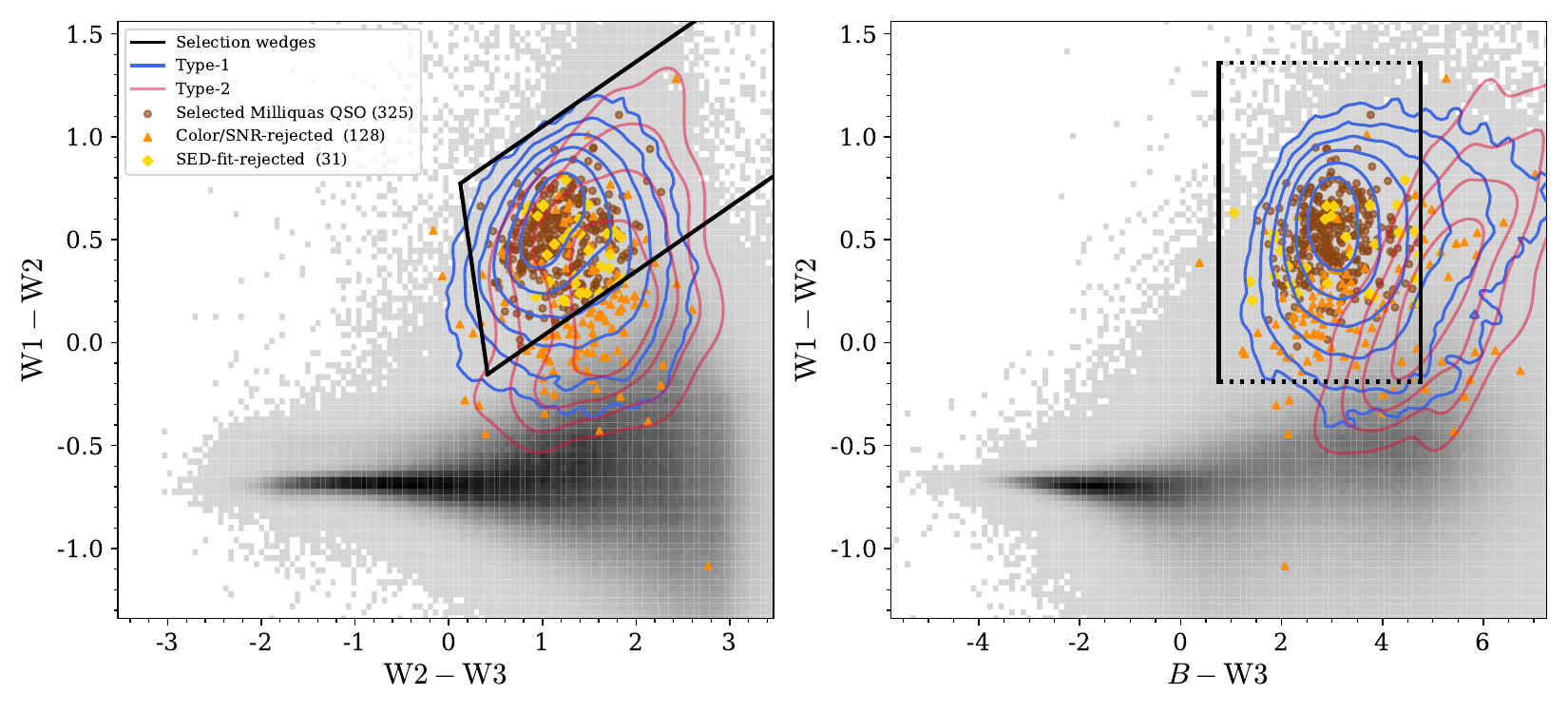}
\caption{Color-color diagrams of the reliable sources that have \texttt{SSFLAG}=Null and \texttt{NDITH}$\geq4$ in the KS4 catalog. The QSO selection criteria are indicated by black lines. The blue contours map the distribution of type-1 QSOs in \cite{Lyke20}, while the red contours represent the distribution of type-2 QSOs \citep{Reyes08}. The blue contour levels range from 0.5$\sigma$ to 3$\sigma$ at 0.5$\sigma$ intervals, starting from the innermost to the outermost, whereas red contour levels range from 0.5$\sigma$ to 2$\sigma$ with the same increment, also moving outward from the center. The squares indicate the spectroscopically confirmed QSOs from the Milliquas v8 catalog \citep{Flesch23} within our survey area. The brown circles represent the QSOs selected by our method, while the orange triangles and yellow diamonds are rejected ones due to their photometry and poor fitting results, respectively.
\label{fig:ccd}}
\end{figure*}

We initially eliminate unreliable sources flagged by the KS4 data processing pipeline. These flagged sources typically arise from spurious signals, such as crosstalk, blended light from bright sources, and instrumental hot/bad pixels (see \texttt{SSFLAG} in the KS4 catalog; S. Chang et al., in preparation). As mentioned in Section \ref{sec:data}, each KS4 field was intentionally dithered at least four times. To ensure data homogeneity, we select objects observed at least four times in every band, using the \texttt{NDITH} flags in the KS4 catalog (\texttt{NDITH\_BVRI}$\geq 4$). Then, the effective survey area size decreases, as shown by the red patches in Figure \ref{fig:coveragemap}.

Our source detection was conducted in the $I$-band images, hence there are numerous sources beyond the 5$\sigma$ imaging depths in the other bands, as shown in Figure \ref{fig:histdepth}.
For reliable source selection, we apply an additional signal-to-noise ratio threshold of 5 in the $I$-band (SNR$_{I} > 5$), corresponding to an $I$-band magnitude cut of $I\lesssim 21$ mag. Note that the SNRs are simply determined by taking the reciprocal of magnitude errors. The imposition of this additional SNR criterion in the $I$-band (shown in red) marginally reduces the number of sources. The $B$-band and WISE detections are also important in the QSO selection below, so we give additional criteria for them: SNR$_{B}>2$, SNR$_{\rm W1}>2$, SNR$_{\rm W2}>2$, and SNR$_{\rm W3}>2$. Further imposing these SNR criteria results in a significant reduction in the source count (shown in blue), while the numbers of bright sources ($I\lesssim16$ mag) are relatively less affected by these criteria. This is due to the remaining bright stars after the selection. Indeed, most of the remaining bright sources have blue MIR colors (${\rm W1}-{\rm W2}\simeq-0.7$), implying that they are not likely to be QSOs.

We distinguish QSOs from stellar objects and inactive galaxies by their MIR colors; QSOs exhibit redder MIR colors due to the hot dust emissions reprocessing light from accretion disks. We adopt the AGN selection criteria based on WISE colors of X-ray-detected sources as proposed by \cite{Mateos12}:

\begin{enumerate}
    \item $(\rm{W1}-\rm{W2}) < 0.315 \times (\rm{W2}-\rm{W3}) + 0.734$
    \item $(\rm{W1}-\rm{W2}) > 0.315 \times (\rm{W2}-\rm{W3}) - 0.284$
    \item $(\rm{W1}-\rm{W2}) > -3.172 \times (\rm{W2}-\rm{W3}) + 1.163$.
\end{enumerate}

\noindent \cite{Mateos12} defined these criteria in the Vega magnitude system. We applied the Vega-to-AB conversion factors as per the AllWISE data \citep{Cutri13}, with a minor discrepancy ($<1\,\%$) from those used in \cite{Mateos12}. Note that most X-ray-detected sources in \cite{Mateos12} are at $z<2$, where their hot dust components are observable within the W1-to-W3 bands. The left panel of Figure \ref{fig:ccd} shows the WISE color-color diagram relevant to the selection wedge (black lines). Accompanying this are the color distributions of spectroscopically confirmed type-1 (blue contours) and type-2 (red contours) QSOs, sourced from \cite{Lyke20} and \cite{Reyes08}, respectively. As this selection wedge is determined by X-ray detection, it effectively captures both type-1 and type-2 QSOs. Our objective is to refine the selection to unobscured QSOs; thus, we impose an additional criterion based on the optical-MIR colors to reflect the bluer optical spectral shape of unobscured QSOs, following \cite{Byun23}:

\begin{enumerate}
  \setcounter{enumi}{3}
  \item $0.755<(B-\rm{W3})\leq 4.755$.
\end{enumerate}

\noindent Note that this criterion is converted to the AB magnitude system, while the previously proposed criterion is given in the Vega magnitude system \citep{Byun23,Byun24}.

The right panel in Figure \ref{fig:ccd} presents the color distributions incorporating $B$-band magnitudes. This fourth criterion, shown as the black solid lines, more accurately differentiates between type-1 and type-2 QSOs. Indeed, 91.5\,\% of the type-1 QSOs from \cite{Lyke20} fit our criteria, whereas a mere 6.2\,\% of the type-2 QSOs from \cite{Reyes08} do so. It is noteworthy that the selection of type-2 QSOs may be skewed by the small sample size (37 out of 595) relative to type-1 QSOs (92k out of 101k). 
We also note that our selection criteria are highly effective for identifying type-1 QSOs up to $z\sim2$, with a recovery rate of 95\,\%. The recovery rate drops significantly to 7\,\% at $z\sim3.3$, indicating that these criteria are primarily suited for the identification of low-redshift type-1 QSOs. This is because high-redshift QSOs exhibit faint flux at the $B$ band and may not be detected in the W3 band.
We further discuss the selection efficiency of spectroscopically confirmed QSOs within the KS4 field in Section \ref{sec:milliquas}.

Consequently, employing the above selection criteria yields 106,443 QSO candidates over the KS4 survey area.

\subsection{SED Fitting} \label{sec:sedfit}

\begin{figure}
\centering
\epsscale{1.2}
\plotone{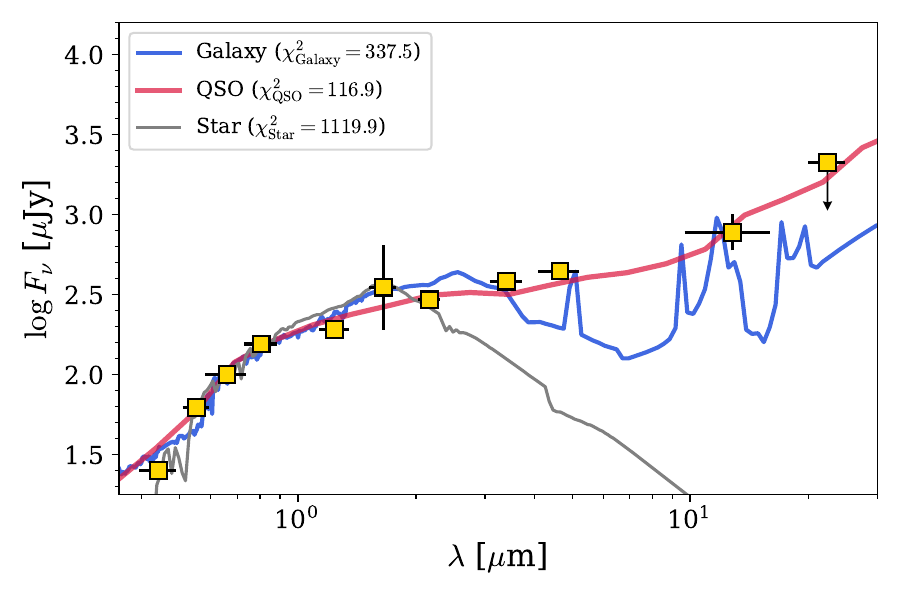}
\caption{Example of SED fitting results. The yellow squares are the photometric data, while the blue, red, and gray lines represent the best-fit galaxy, QSO, and star models, respectively.  The $\chi^{2}$ values of the models are marked in the legend.
\label{fig:sed}}
\end{figure}

Because the photometric selection solely with broad-band photometry can lead to including a large portion of non-QSOs, we carry out the SED fitting on the pre-selected sources in order to further perform the rigorous classification. The detailed method is described in \cite{Son23} and \cite{Byun23}. The photometric data obtained from KS4, 2MASS, VHS, and WISE is used for the analysis. We adopt LePhare\texttt{++}\footnote{\url{https://gitlab.lam.fr/Galaxies/LEPHARE}}, a C\texttt{++} rendition of the original Fortran program LePhare \citep{Arnouts99,Ilbert06}, which allows us to fit the SED with various sets of SED templates. We initially employ three AGN templates from \cite{Lyu17}, which were generated based on the dust features in mid-infrared: normal, hot dust-deficient, and warm dust-deficient AGNs. In addition, we consider the emission from the polar dust with extinction ($A_v = 0-1.25$ mag) on the AGN continuum. Finally, the contribution from the host galaxies, which is modeled with 7 types of SWIRE galaxies \citep{Polletta07} (old stellar population with an age of 7 Gyr, and Hubble types of E, S0, Sa, Sb, Sc, Sd) is added. The flux ratio of the host to the total flux at 1.6 $\mu$m is set to be $1-95$\%. Finally, 32 SED templates of normal galaxies used for photo-z estimation of the COSMOS survey \citep{Ilbert09} and the stellar templates from \cite{Bohlin95} and  \cite{Pickles98} are adopted. Figure \ref{fig:sed} shows an example of the SED fitting results.

To select the best model among the SED templates for QSOs, inactive galaxies, and stars in describing the observed SEDs of the QSO candidates, we employ the Bayesian information criterion (BIC). The BIC is defined as ${\rm BIC}=\chi^2 + k \ln{n}$, where $k$ is the number of free parameters and $n$ is the number of the data points in the SED. According to this criterion, a source is classified as a QSO only if the BIC for the best fit with QSO templates is at least 10 points lower than that of the best fit with galaxy SED templates \citep{Little07}, a method proven effective for QSO/galaxy classification \citep{Shin20,Byun23}. Additionally, to discard heavily obscured AGNs, we impose an additional criterion on extinction ($\tau_\nu \leq 1$). We further refine our selection by excluding sources whose $\chi^{2}_{\rm QSO}$ values exceed the 97.7\,\% (2$\sigma$) confidence threshold. This statistical cutoff helps identify and remove outliers or less likely QSO candidates based on the quality of their fit, ensuring that only the most probable QSOs are retained in our analysis.  Employing these stringent criteria enhances the integrity and reliability of our dataset, focusing on sources that best match the expected characteristics of QSOs.

In the KS4 field, the final set of unobscured QSO candidates comprises 72,964 sources, all of which meet the selection criteria outlined in Sections \ref{sec:colorsel} and \ref{sec:sedfit}. We provide a detailed multi-wavelength catalog of these candidates, as described in Table \ref{tbl:catalog}.

\begin{deluxetable}{lccl}
\tabletypesize{\scriptsize}
\tablewidth{0pt}
\tablecaption{Description of the Columns in the KS4 QSO Candidate Catalog. \label{tbl:catalog}}
\tablehead{
\colhead{Column} & \colhead{Format} & \colhead{Unit} & \colhead{Description}
}
\startdata
ID & String & & KS4 QSO candidate designation\\
RA & Double & deg & R.A. (J2000) \\
DEC & Double & deg & Decl. (J2000) \\
\hline
Bmag & Double & mag & KS4 $B$-band MAG\_AUTO magnitude \\
e\_Bmag & Double & mag & Error on $B$-band magnitude\\
Vmag & Double & mag & KS4 $V$-band MAG\_AUTO magnitude \\
e\_Vmag & Double & mag & Error on $V$-band magnitude\\
Rmag & Double & mag & KS4 $R$-band MAG\_AUTO magnitude \\
e\_Rmag & Double & mag & Error on $R$-band magnitude\\
Imag & Double & mag & KS4 $I$-band MAG\_AUTO magnitude \\
e\_Imag & Double & mag & Error on $I$-band magnitude\\
\hline
Jmag & Double & mag & $J$-band magnitude \\
e\_Jmag & Double & mag & Error in $J$-band \\
f\_Jmag$^{\dagger}$ & Int & & Flag in $J$-band  \\
Hmag & Double & mag & $H$-band magnitude \\
e\_Hmag & Double & mag & Error in $H$-band \\
f\_Hmag$^{\dagger}$ & Int & & Flag in $H$-band \\
Kmag & Double & mag & $K_s$-band magnitude \\
e\_Kmag & Double & mag & Error in $K_s$-band \\
f\_Kmag$^{\dagger}$ & Int & & Flag in $K_s$-band \\
\hline
W1mag & Double & mag & WISE W1-band magnitude \\
e\_W1mag & Double & mag & Error on W1-band magnitude \\
W2mag & Double & mag & WISE W2-band magnitude \\
e\_W2mag & Double & mag & Error on W2-band magnitude \\
W3mag & Double & mag & WISE W3-band magnitude \\
e\_W3mag & Double & mag & Error on W3-band magnitude \\
W4mag & Double & mag & WISE W4-band magnitude \\
e\_W4mag & Double & mag & Error on W4-band magnitude \\
\enddata
\tablecomments{All the magnitudes are given in AB magnitudes. Unavailable values are indicated as $-99.0$. A magnitude with an error of $-99.0$ means an upper limit. (This table including the full data is published in its entirety in the machine-readable format.) }
\tablenotetext{\dagger}{0: no data, 1: VHS, 2: 2MASS}
\end{deluxetable}

\section{Catalog Validation} \label{sec:validation}

The most direct method to confirm our QSO candidates as real QSOs is to obtain their optical/near-infrared spectra to identify the AGN features. However, confirming all QSO candidates through individual spectral observations is a cost-ineffective approach for conducting QSO surveys in the southern sky. We, therefore, anticipate relying on forthcoming spectro-photometric surveys, such as 7DS and SPHEREx, for their direct confirmation. Instead, we here explore the validity of our QSO candidates through indirect methods.

\subsection{Cross-match with Known QSOs \label{sec:milliquas}}

\begin{figure*}
\centering
\epsscale{1.1}
\plottwo{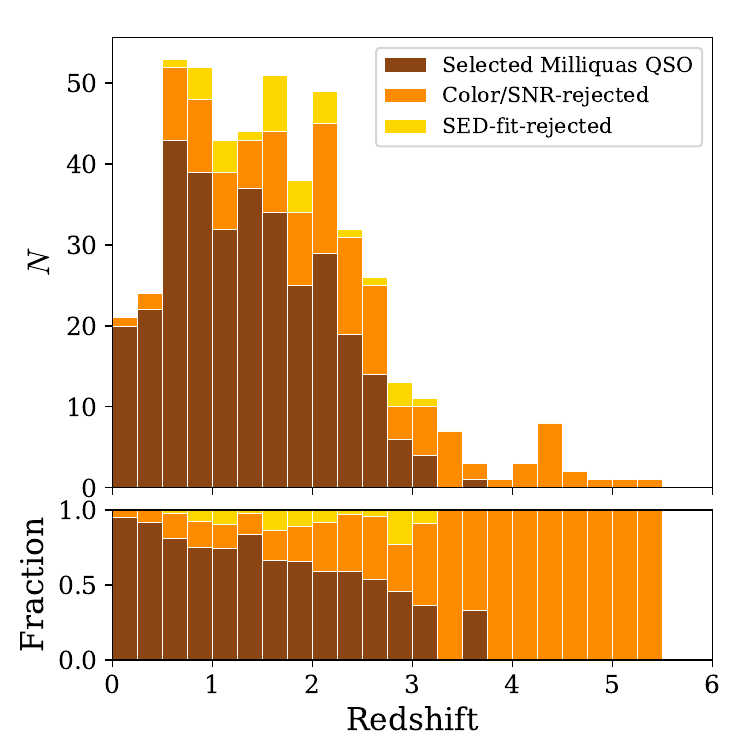}{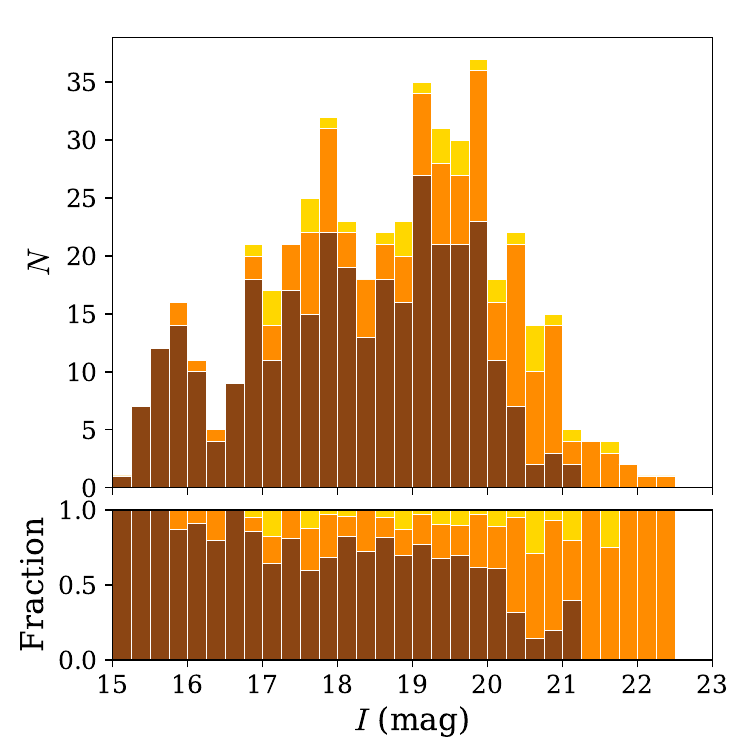}
\caption{Redshift (left) and $I$-band magnitude (right) distributions of Milliquas QSOs within the KS4 survey area. The top and bottom panels show the stacked histograms and the fractional distributions, respectively. The brown histograms represent the QSOs that meet our selection criteria, while the orange and yellow histograms are the rejected ones because of their photometry and poor fitting results, respectively.
\label{fig:mq_zmag}}
\end{figure*}

To validate our selection, we compare our QSO candidates with spectroscopically confirmed QSOs in the KS4 survey area. We utilize the Milliquas catalog v8 \citep{Flesch23}, which contains approximately 0.9 million type-1 QSOs across the entire sky, with only a few located within our survey area. Specifically, within our survey area, there are 484 QSOs classified as the core-dominant type-1 QSOs in the catalog. Upon cross-matching with our QSO candidates, we find only 325 QSOs, resulting in a recovery rate of 67\,\%.

In Figure \ref{fig:ccd}, we present the color distributions of the 484 QSOs from the Milliquas catalog, having reliable KS4 or WISE photometry. Approximately a quarter of QSOs are rejected during the color-selection stage (orange squares), most of which are situated just beneath the selection wedge of \cite{Mateos12} in the left panel, a region where only a very minor fraction of type-1 QSOs are expected to be located (blue contours). The remaining samples are rejected during the selection based on the SED fitting (yellow squares). It is important to note that the Milliquas sample is a compilation of spectroscopically confirmed QSOs without homogenetic selection, which may be related to the missing populations by our method.

In Figure \ref{fig:mq_zmag}, we present the distributions of redshift (left) and $I$-band magnitude (right) for the Milliquas QSOs within the KS4 survey area. Since our selection criteria are optimized for identifying low-redshift, unobscured QSOs, the missing fraction of QSOs increases at higher redshifts. This fraction also increases at $I\gtrsim20$ mag, where the number count of reliable sources after applying the SNR criteria drops dramatically (Figure \ref{fig:histdepth}). As pointed out in Section \ref{sec:colorsel}, our color-selection criteria are highly effective for finding low-redshift QSOs. If we consider the QSOs with $I<20$ mag at $z<2$, the numbers of selected, color/SNR-rejected, and SED-fit-rejected QSOs are 232, 21, and 13, respectively, resulting in a recovery rate of 87\,\% (232/266). It is noteworthy that the fraction of color-selected objects is then 92\,\% [(232+13)/266], comparable to the selection efficiency for SDSS type-1 QSOs as discussed in Section \ref{sec:colorsel}. This similarity highlights the effectiveness of our QSO selection methods for spectroscopically confirmed QSOs in the southern sky.

\subsection{Number Counts} \label{sec:number}

\begin{figure*}[t]
\centering
\epsscale{1.1}
\plotone{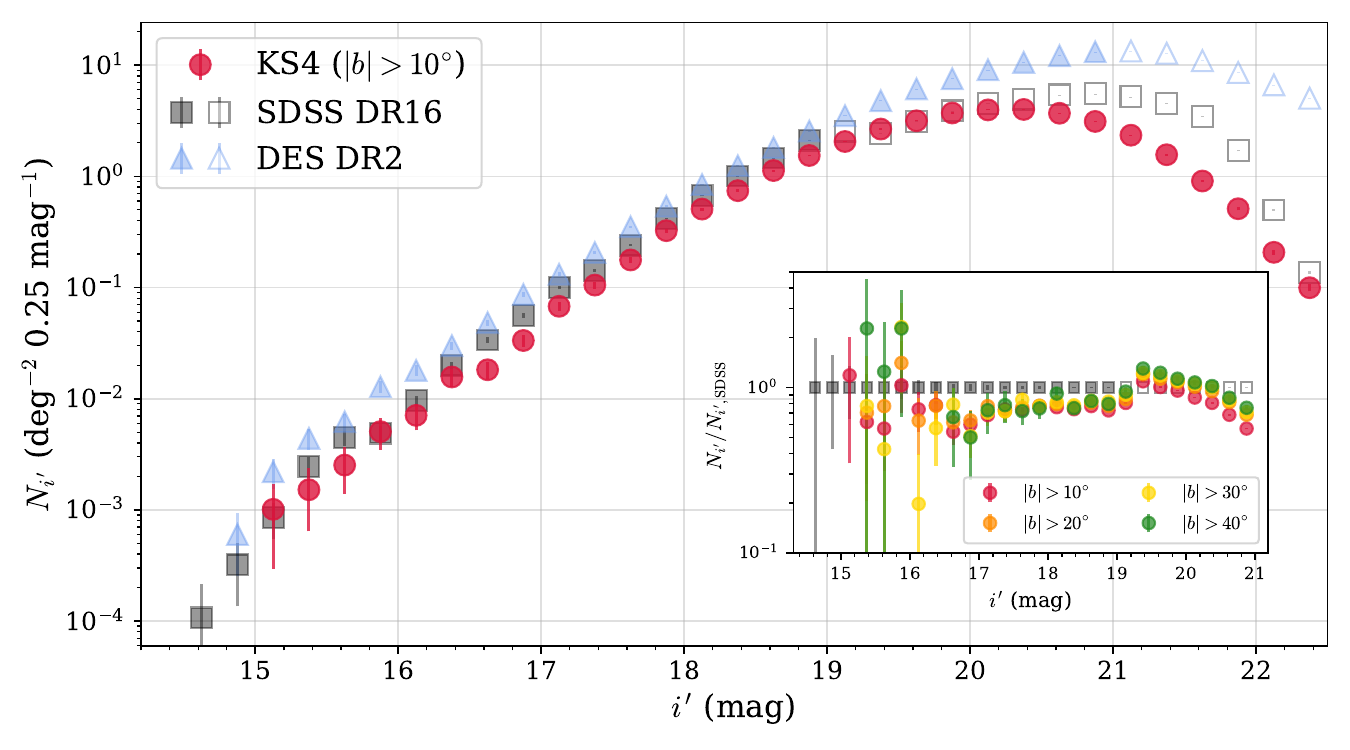}
\caption{Differential number counts of KS4 QSO candidates in the $i'$-band (red circles) at $|b|>10^{\circ}$. The gray squares represent spectroscopically identified unobscured QSOs in SDSS \citep{Lyke20}, while the open symbols indicate incomplete bins. Similarly, the filled and open sky-blue triangles are for the DES DR2 QSO candidates \citep{Yang23}. The inset shows the number counts of KS4 QSO candidates with varying galactic latitude limitations, compared to those of SDSS QSOs.
\label{fig:nc}}
\end{figure*}

As an indirect approach to validate our QSO candidates, we compare their number counts per unit area to those of spectroscopically identified QSOs from the SDSS DR16 \citep{Lyke20}, spanning an area of 9376\,deg$^{2}$. 
Furthermore, we restrict our analysis to QSOs with a redshift lower than 2 to accommodate the variations in number counts by redshift. The adjusted number counts from SDSS are represented as gray squares in Figure \ref{fig:nc}. Note that the SDSS QSO sample is incomplete at $i'>19.1$ mag \citep{Vanden05}, indicated by open squares in the figure.


To match the number counts of SDSS QSOs, we convert KS4 magnitudes of our QSO candidates to SDSS $i'$-band magnitude using the combination of transformation equations for $z\leq2.1$ QSOs by \cite{Jester05}: 
$i' = V -0.19(B-V)-0.9(R-I)+0.18$.

\begin{figure}
\centering
\epsscale{1.1}
\plotone{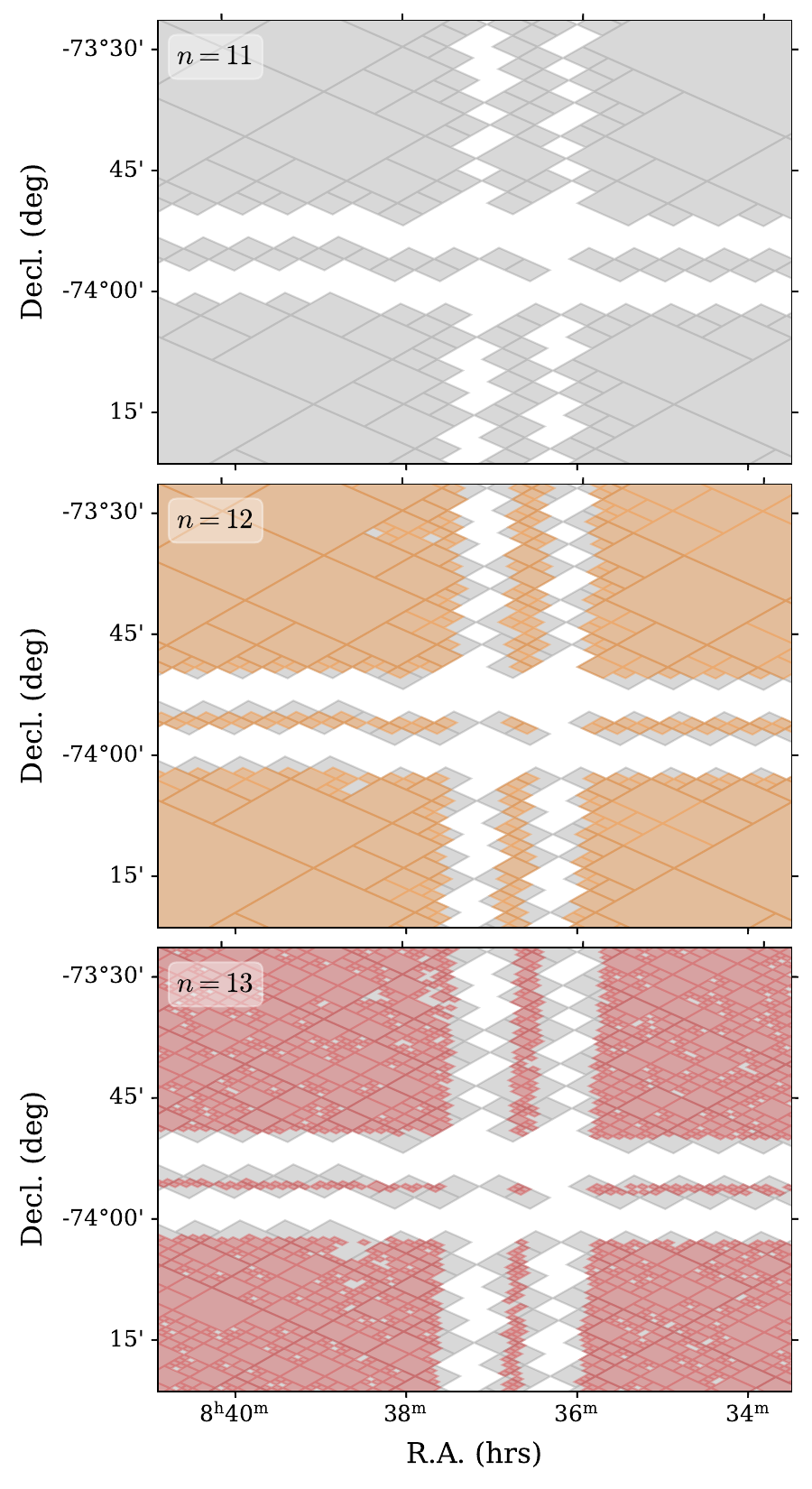}
\caption{Changes in sky coverage depending on HEALPix Levels of 11 (gray), 12 (orange), and 13 (red). Each panel presents the effective area coverage for a given HEALPix level, with areas from the $n=11$ level superimposed to enable direct comparison. The unoccupied regions are attributable to the gaps between the KMTNet CCD chips, representing incomplete regions not covered by at least four dithered observations or the areas affected by bright stars.
\label{fig:area}}
\end{figure}

\begin{deluxetable}{ccccc}
\tabletypesize{\scriptsize}
\tablewidth{0pt}
\tablecaption{Effective Survey Area \label{tbl:area}}
\tablehead{
\colhead{} & \multicolumn{3}{c}{HEALPix levels} & \colhead{$N$}\\
\cmidrule(lr){2-4} 
\colhead{$|b|$} & \colhead{11} & \colhead{12} & \colhead{13} &
}
\startdata
$>10^{\circ}$ & 2077.2 & 1977.9 & 1731.1 & 72,636\\
$>20^{\circ}$  & 1217.2 & 1158.1 & 975.5  & 48,924\\
$>30^{\circ}$  & 555.7 & 527.3 & 430.0 & 23,090\\
$>40^{\circ}$  & 188.9 & 179.1 & 141.2 & 8147\\
\enddata
\tablecomments{The units of $b$ and area are deg and deg$^{2}$, respectively. The last column is for the number of QSO candidates.}
\end{deluxetable}

Accurately determining the effective survey area is crucial for precise estimation of the number counts. Considering the gaps between KMTNet CCD chips carefully, we introduce the Hierarchical Equal Area isoLatitude Pixelation of a sphere (HEALPix; \citealt{Gorski05}), which divides the sphere's surface into uniform-sized areas according to given levels. We count the HEALPix pixels covering the reliable sources in the KS4 catalog (\texttt{SSFLAG}$=$Null \& \texttt{NDITH}$\,\geq 4$). Figure \ref{fig:area} shows examples of the source-matched pixels depending on HEALPix levels ranging from 11 to 13, corresponding to spatial resolutions of 2.95, 0.74, and 0.18 arcmin$^{2}$, respectively. There are gaps between the CCD chips, which are not covered by at least four dithered observations. As the level goes higher, the polygons trace the shape of the gap sharply, while the unoccupied regions by bright stars appear more as well. However, at higher HEALPix levels, the finer resolution results in some pixels being excluded from the calculation due to their location in regions with no detectable sources, even though these regions are covered more than four times (\texttt{NDITH}$\geq4$). This exclusion leads to an underestimation of the survey area. Consequently, we adopt the effective area under the assumption of HEALPix level 12. The sizes of effective survey areas according to galactic latitude ($b$) limit are listed in Table \ref{tbl:area}.

The red circles in Figure \ref{fig:nc} represent the differential number counts of KS4 QSO candidates located at $|b|>10^{\circ}$. While the number counts from our candidates generally align with those observed in the SDSS QSO sample, they are marginally lower in the complete bins ($i'< 19.1$ mag). This discrepancy may be attributed to several factors: (1) a potentially missing population not captured by our selection criteria, (2) the uncertainties associated with the estimation of the SDSS survey area size (9376\,deg$^{2}$; \citealt{Paris18}), and (3) the omission of precise corrections for selection completeness, which approaches unity ($>90\,\%$) at $i\lesssim19$ mag.  Nevertheless, the power-law slopes of 0.85 for KS4 and 0.83 for SDSS within the magnitude range of $16<i'<19$ suggest that our selection method effectively reproduces the observed abundance of real QSOs.

The inset compares the number counts of our candidates, defined by various $b$ limits, normalized to those of the SDSS QSOs. This comparison shows consistent trends across different $b$ thresholds, further validating the efficacy of our selection criteria. In particular, no noticeable number excess between $|b|>10^{\circ}$ and $|b|>40^{\circ}$ suggests that our method effectively excludes stellar contaminants.

\cite{Yang23} recently published a photometric catalog of QSO candidates from the DES Data Release 2,  covering $\sim5000$\,deg$^{2}$ of the southern sky. This catalog utilizes optical-to-IR photometry via a probabilistic approach to identify candidates. The number counts from this sample, represented by sky-blue triangles in Figure \ref{fig:nc}, are about twice as high as those from our candidates. Note that the sample used here adheres to the higher-purity recommendations (criteria (1)--(6) in \citealt{Yang23}) and considers QSOs with photometric redshifts lower than 2. Similar to our comparison with the SDSS sample, the noted discrepancy could stem from the previously mentioned factors. Additionally, differences may also arise from the inclusion of host-dominant AGNs from the Miliquas catalog in the training sample, which our selection criteria may largely overlook. The power-law slope between $16<i'<19$ is slightly flatter at $0.78$, potentially reflecting the broader inclusivity of their targeting strategy.

We further emphasize the complementarity in sky coverage between the KS4 QSO catalog and the DES DR2 catalog of \cite{Yang23}. These two catalogs cover most of the southern sky (see Figure \ref{fig:coveragemap}), providing an extensive resource for QSO studies.


\subsection{Crosscheck with Gaia DR3}

The Gaia mission is designed to observe bright sources ($G \leq 20.7$ mag in the Vega system) in the all-sky \citep{Gaia16}. In the recent data release 3 (DR3; \citealt{Gaia23}), they present the probability of being a QSO ($P_{\rm QSO}$) determined by the discrete source classifier (DSC). The Combmod probability is determined from the combination of class information from Specmod (based on spectra) and Allosmod (based on photometric and astrometric data). Similarly, they provide the probabilities of being a galaxy ($P_{\rm Gal}$) and a star ($P_{\rm Star}$).

\begin{figure}
\centering
\epsscale{1.1}
\plotone{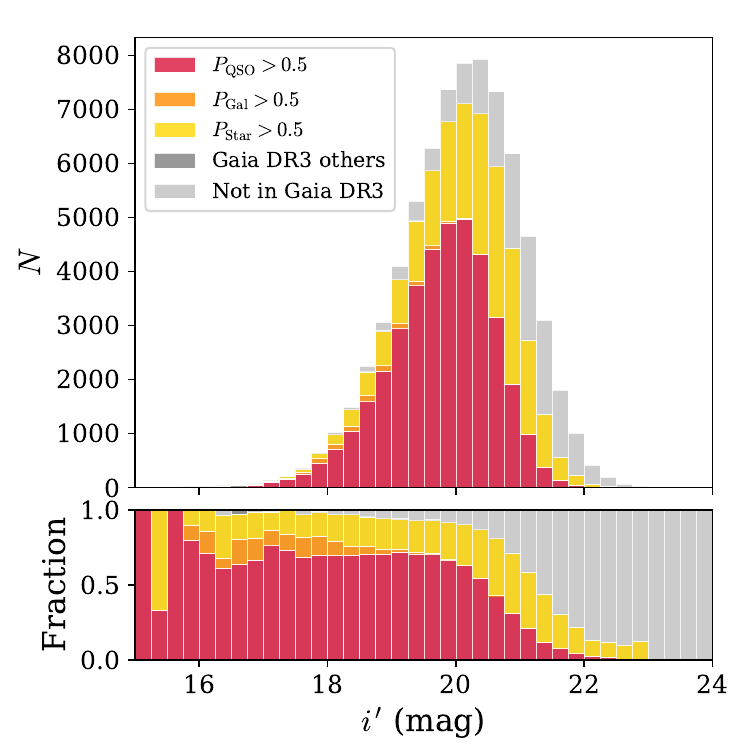}
\caption{Stacked histogram (top) and fractional distribution (bottom) of our QSO candidates, with classifications based on inclusion in the Gaia DR3 dataset. The colors represent different classifications based on probability thresholds: red for $P_{\rm QSO}>0.5$, orange for $P_{\rm Gal}>0.5$, and yellow for $P_{\rm Star}>0.5$. Gray indicates those within Gaia DR3 that do not meet any of the specified probability criteria, while light-gray represents candidates not matched to Gaia DR3 sources.
\label{fig:pgaia}}
\end{figure}

In Figure \ref{fig:pgaia}, we present the fractional distribution of our candidates as classified by the probabilities in Gaia DR3. According to Gaia DR3's source classification, sources are deemed QSOs, galaxies, or stars based on the highest posterior probability exceeding 0.5. Notably, only 64\,\% of our candidates are likely classified as QSOs with $P_{\rm QSO}>0.5$ (red histogram), while a significant portion (34\,\%) of the remaining sources are likely classified as stars (yellow histogram). 

This classification disparity appears to contradict the high completeness reported for QSO and star classifications using Combmod (0.916 and 0.996, respectively; \citealt{Ulla22}). However, it is important to note that these probabilities were determined using only Gaia data in optical wavelengths. Even if an object is a point source with $P_{\rm Star}>0.5$, the presence of detection in the WISE bands suggests a significant possibility that the object might not be a star, except for very bright sources ($I\lesssim16$ mag). This claim is supported by the star SED shown in Figure \ref{fig:sed} and by the changes in the number distribution after the SNR cut in the WISE bands due to the inclusion of bright stars, shown in Figure \ref{fig:histdepth}. Furthermore, cross-matching with the Milliquas catalog in Section \ref{sec:milliquas} reveals that 20\,\% of spectroscopically identified QSOs are assigned $P_{\rm Star}>0.5$.  Additionally, the effectiveness of our approach in rejecting stellar contamination, strengthened by the consistent trends in the number counts across different $b$ thresholds (Figure \ref{fig:nc}), suggests that not all sources with $P_{\rm Star}>0.5$ in our candidates are indeed stars. These raise concerns about the reliability of using $P_{\rm QSO}$ alone for accurate QSO selection.

\begin{figure}
\centering
\epsscale{1.1}
\plotone{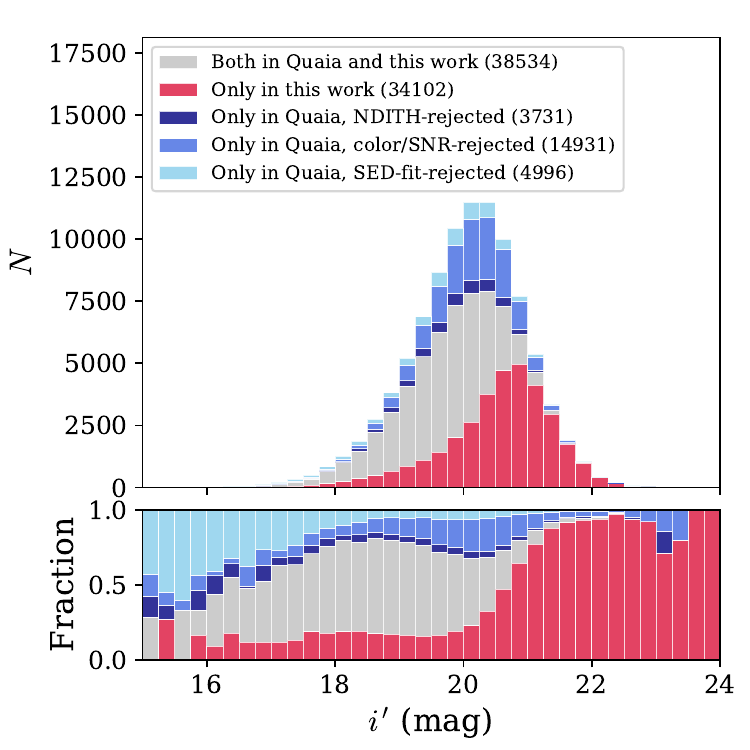}
\caption{Stacked histogram (top) and fractional distribution (bottom) of QSO candidates at $|b|>10^{\circ}$ selected in Quaia and this work. The gray histogram represents the candidates both in Quaia and this work. The red histogram indicates the candidates only in this work. The navy, blue, and sky-blue histograms are those rejected due to \texttt{NDITH}, color/SNR criteria, and SED-fit results, respectively.
\label{fig:quaia}}
\end{figure}

Given doubts about the low purity (0.240) of the Gaia DR3 QSO candidates with $P_{\rm QSO}$ selection \citep{Ulla22}, \cite{Storey-Fisher24} recently generated an all-sky QSO catalog named Quaia, based on the combination of the Gaia DR3 and unWISE \citep{Lang14} colors, yielding a more reliable set of QSO candidates. There are 64,734 Quaia candidates located within the effective survey area defined by HEALPix level $n=12$, 2542 of which are not identified in the KS4 catalog. This omission may arise from the marginal inconsistency between the imaging data and HEALPix patches. Indeed, about 6\,\% of the Quaia candidates matched to the KS4 catalog (3731/62,192) do not satisfy the \texttt{NDITH} criteria used for the effective survey area estimation. The fraction naturally decreases if we introduce an effective survey area defined by a higher HEALPix level (e.g., $\sim4$\,\% with $n=13$). We also note that the screening with \texttt{SSFLAG} has no effect because it excludes known sources.

In Figure \ref{fig:quaia}, we present the $i'$-band distribution of these Quaia candidates matched to the KS4 catalog and our sample, considering the 38,534 candidates selected both in Quaia and our work (gray histogram). On the other hand, 23,593 Quaia candidates (38\,\%) are not selected by our selection criteria; they are rejected due to (1) the lack of the number of dithered observations (\texttt{NDITH}$\geq4$; navy), (2) our photometric selection criteria (SNR and colors; blue), and (3) poor SED fitting results (sky-blue). The \texttt{NDITH}-rejected candidates are because we select Quaia candidates using the KS4 HEALPix maps, which cannot perfectly trace our survey area, especially at the CCD gaps. Unlike our method, \cite{Storey-Fisher24} used only the W1 and W2 band magnitudes from the unWISE catalog, which is deeper than the AllWISE that we used, potentially explaining the discrepancies in candidate selection due to SNR and colors. Indeed, the magnitude differences in W1 and W2 between unWISE\footnote{We used the \texttt{astroquery} Python package \citep{Ginsburg19} to find the matched sources in the unWISE catalog \citep{Lang14}.} and AllWISE for the Quaia candidates excluded based on color/SNR exhibit larger standard deviation (0.41 mag for W1, 0.28 mag for W2) compared to those candidates selected both in Quaia and in our analysis (0.16 mag for W1, 0.13 mag for W2). This may also be attributed to either the MIR variability between the two surveys (e.g., \citealt{Son22}) or the confusion with nearby sources. On the other hand, those rejected due to the poor SED-fitting results occupy a large fraction of bright QSO candidates ($i'<18$ mag). Most of them have $\chi^{2}_{\rm QSO}$ values higher than $2\sigma$ level, indicating that they are unlikely to be probable QSOs or nearby QSOs with the bright and extended host galaxies, in which the systematic uncertainties in the multi-wavelength photometry can be significantly larger than nucleus dominated objects. 

The number of QSO candidates only in our work (red histogram) is significantly high, predominantly consisting of faint sources that are not likely to be observable with Gaia. However, even at $i'\lesssim20$ mag, the fraction of these candidates remains nonnegligible; about 17\,\% at $15 < i' < 20$. As we discussed above, the selection efficiency of our method, validated by the recovery rate of spectroscopically confirmed QSOs, and the number counts consistent with the other surveys, strengthens the fact that our QSO candidates are promising. Therefore, spectro-photometric surveys, such as 7DS and SPHEREx, will enable us to estimate the effectiveness of our selection method rigorously and to constrain the unobscured QSO population in the southern sky.

\section{Summary}

In this study, we present a catalog of unobscured QSO candidates in the southern sky. We mainly use the KS4 interim data, which covers $\sim 2500$\, deg$^{2}$ area around the south ecliptic pole and achieves 5$\sigma$ imaging depths of $\sim22.1$--22.7 mag in the $BVRI$ bands. Combining this KS4 data with infrared photometric data from the 2MASS ($JHK_s$), VHS ($JHK_s$), and AllWISE (W1W2W3W4) surveys, we perform the color selection for the initial selection and apply the SED fitting to refine our list of plausible QSO candidates. The final catalog consists of 72,964 candidates for unobscured QSOs over an effective survey area of $\sim2000$\,deg$^{2}$. Despite only 0.4\,\% of these candidates being spectroscopically confirmed QSOs so far, the high recovery rate of 87\,\% for QSOs with $I<20$ mag at $z<2$ proves the robustness of our selection method. Moreover, this is also supported by the number counts of our candidates, which are consistent with those of the spectroscopically confirmed QSOs from SDSS in the northern hemisphere. Moving forward, upcoming spectro-photometric surveys, such as SPHEREx and 7DS, are expected to provide valuable insights into the true nature of these candidates, thereby enhancing our understanding of QSO populations in the southern sky.


\section{Acknowledgments}

We thank the anonymous referee for valuable suggestions that greatly improved the manuscript. This work was supported by the National Research Foundation of Korea (NRF) grant funded by the Korean government (MSIT) (No. 2021R1C1C2091550, 2022R1A4A3031306, RS-2024-00347548, 2021R1C1C1013580). 

MI, SWC, JMK acknowledges support from the National Research Foundation of Korea (NRF) grants, No. 2020R1A2C3011091 and No. 2021M3F7A1084525 funded by the Ministry of Science and ICT (MSIT). This research was also supported by Basic Science Research Program through the NRF funded by the Ministry of Education (RS-2023-00245013).

This research has made use of the KMTNet system operated by the Korea Astronomy and Space Science Institute (KASI) at three host sites of CTIO in Chile, SAAO in South Africa, and SSO in Australia.
Data transfer from the host site to KASI was supported by the Korea Research Environment Open NETwork (KREONET).

This publication makes use of data products from the Wide-field Infrared Survey Explorer, which is a joint project of the University of California, Los Angeles, and the Jet Propulsion Laboratory/California Institute of Technology, and NEOWISE, which is a project of the Jet Propulsion Laboratory/California Institute of Technology. WISE and NEOWISE are funded by the National Aeronautics and Space Administration.

This publication makes use of data products from the Two Micron All Sky Survey, which is a joint project of the University of Massachusetts and the Infrared Processing and Analysis Center/California Institute of Technology, funded by the National Aeronautics and Space Administration and the National Science Foundation.

The VISTA Hemisphere Survey data products served at Astro Data Lab are based on observations collected at the European Organisation for Astronomical Research in the Southern Hemisphere under ESO programme 179.A-2010, and/or data products created thereof.

The national facility capability for SkyMapper has been funded through ARC LIEF grant LE130100104 from the Australian Research Council, awarded to the University of Sydney, the Australian National University, Swinburne University of Technology, the University of Queensland, the University of Western Australia, the University of Melbourne, Curtin University of Technology, Monash University and the Australian Astronomical Observatory. SkyMapper is owned and operated by The Australian National University's Research School of Astronomy and Astrophysics. The survey data were processed and provided by the SkyMapper Team at ANU. The SkyMapper node of the All-Sky Virtual Observatory (ASVO) is hosted at the National Computational Infrastructure (NCI). Development and support of the SkyMapper node of the ASVO has been funded in part by Astronomy Australia Limited (AAL) and the Australian Government through the Commonwealth's Education Investment Fund (EIF) and National Collaborative Research Infrastructure Strategy (NCRIS), particularly the National eResearch Collaboration Tools and Resources (NeCTAR) and the Australian National Data Service Projects (ANDS).

This work has made use of data from the European Space Agency (ESA) mission {\it Gaia} (\url{https://www.cosmos.esa.int/gaia}), processed by the {\it Gaia} Data Processing and Analysis Consortium (DPAC, \url{https://www.cosmos.esa.int/web/gaia/dpac/consortium}). Funding for the DPAC has been provided by national institutions, in particular the institutions participating in the {\it Gaia} Multilateral Agreement.

%

\facilities{KMTNet, IRSA}


\software{astropy \citep{astropy13,astropy18,astropy22},
LePhare\texttt{++} (\url{https://gitlab.lam.fr/Galaxies/LEPHARE}), 
Source Extractor \citep{Bertin96}, SCAMP \citep{Bertin2006ASPC..351..112B}, SWarp \citep{2010ascl.soft10068B}, 
Aladin \citep{Bonnarel00}, sfdmap (\url{https://github.com/kbarbary/sfdmap}), astroquery \citep{Ginsburg19}
}





\bibliography{ref}{}
\bibliographystyle{aasjournal}






\end{document}